# A statistical learning framework for mapping indirect measurements of ergodic systems to emergent properties

Nicholas Hindley,[1,2] Stephen J. DeVience,[3] Ella Zhang,[1] Leo L. Cheng,[1,4] Matthew S. Rosen[1,4,5]*

[1] Athinoula A. Martinos Center for Biomedical Engineering, Massachusetts General Hospital, Charlestown, MA 02129, USA
[2] Image X Institute, University of Sydney, Sydney, NSW, Australia
[3] Scalar Magnetics, LLC, Ellicott City, MD 02143, USA
[4] Harvard Medical School, Boston, MA 02115, USA
[5] Department of Physics, Harvard University, Cambridge, MA 02138, USA
* msrosen@mgh.harvard.edu

**Abstract**

The discovery of novel experimental techniques often lags behind contemporary theoretical understanding. In particular, it can be difficult to establish appropriate measurement protocols without analytic descriptions of the underlying system-of-interest. Here we propose a statistical learning framework that avoids the need for such descriptions for ergodic systems. We validate this framework by using Monte Carlo simulation and deep neural networks to learn a mapping between low-field nuclear magnetic resonance spectra and proton exchange rates in ethanol-water mixtures. We found that trained networks exhibited normalized-root-mean-square errors of less than 1% for exchange rates under 150 s$^{-1}$ but performed poorly for rates above this range. This differential performance occurred because low-field measurements are indistinguishable from one another at fast exchange. Nonetheless, where a discoverable relationship between indirect measurements and emergent dynamics exists, we demonstrate the possibility of approximating it without the need for precise analytic descriptions, allowing experimental science to flourish in the midst of ongoing theoretical work.

## Introduction

In experimental science we often seek to measure stable, emergent properties of complex, stochastic processes. For instance, the temperature of a room emerges out of the random thermal motion of particles. In particular, temperature reflects the average kinetic energy of these particles and can be measured by a thermometer, but this measurement procedure probes the underlying stochastic behavior indirectly (e.g. by volume changes of a fluid or the emission of thermal radiation). Therefore, simple measurements can reflect a complex interplay between unobserved stochastic interactions, emergent behavior and a third observable property that links the two (Fig. 1). In the measurement of temperature, this interplay is mediated by precise analytic descriptions (e.g. via gas laws). In fact, the SI unit of temperature, the kelvin, was redefined as recently as 2019 by setting an exact numerical value for the Boltzmann constant (1). Indeed, the meter, the ampere and the mole were similarly redefined in the same year to reflect a shift away from using tangible objects to define fundamental physical quantities toward analytic definitions based on mathematical derivation. This indicates a nexus between theory and measurement where the development of novel experimental techniques, often by necessity, lags behind the discovery of analytic descriptions. Here we propose a statistical learning framework that avoids the need for such descriptions for a particular class of stochastic systems – namely, *ergodic* systems – thereby allowing experimental science to flourish in the midst of ongoing theoretical work.

Initially introduced by Boltzmann while exploring the kinetic theory of gases, the ergodic hypothesis states that all accessible microstates are occupied with equal probability over a long enough time horizon (2). Further developed by Birkhoff (3, 4) and von Neumann (5, 6), ergodic theory has become a mainstay of statistical mechanics and provides mathematical grounding for common-sense notions of randomness. An important property of ergodic systems is that the time-average equals the space-average. That is, for a measure-preserving transformation $T: X \to X$ on some measure space $(X, \Sigma, \mu)$ with μ-integrable function $f$, $\lim_{n \to \infty} \frac{1}{n} \sum_{k=0}^{n-1} f(T^k x) = \frac{1}{\mu(X)} \int f \, d\mu$ almost everywhere. Hence, studying an appropriate cross-section of an ergodic system across space suffices to infer emergent dynamics over time and *vice versa*. Here we leverage this property to learn a mapping between indirect measurements of ergodic processes and emergent behavior. More specifically, we use neural networks to learn such mappings from training data generated via Monte Carlo simulation.

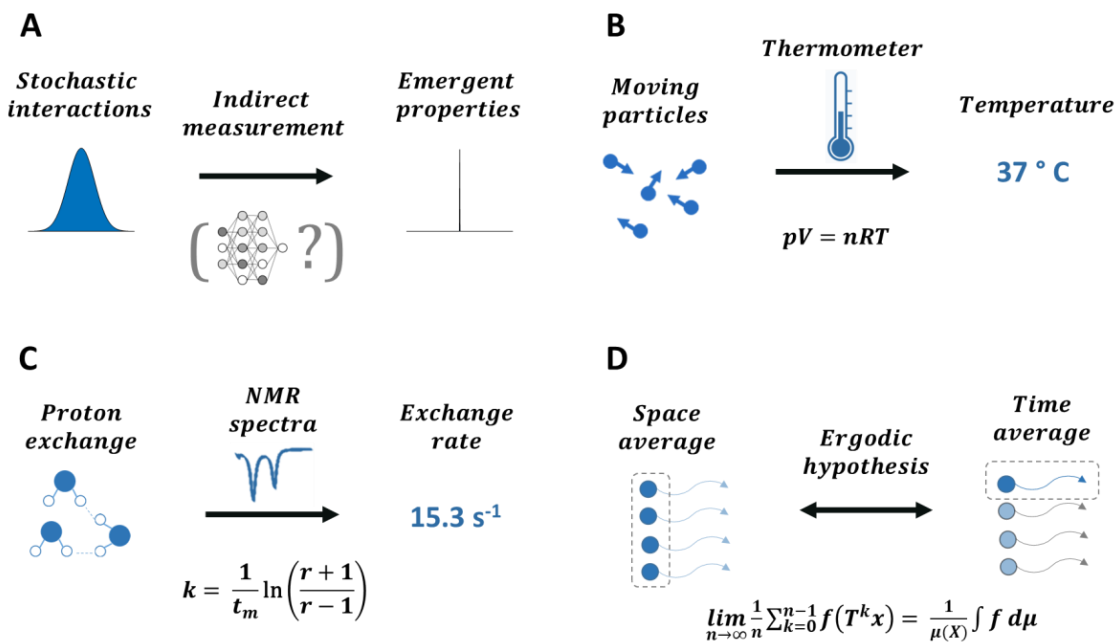

**Fig. 1. Indirect measurements can be used to deduce emergent properties** (**A**) "Fuzzy" stochastic processes can be mapped to "sharp" emergent properties via indirect measurements. (**B**) A familiar example involves the use of a thermometer to measure temperatures which arise due to random thermal motion. With a gas thermometer, temperature can be calculated analytically using the ideal gas law (**C**) Here we explore the link between stochastic proton exchange and the emergent exchange rate via NMR spectroscopy. In high-field NMR, the exchange rate can be calculated analytically using the intensities of proton signals. (**D**) Ergodic systems have the special property that the space-average equals the time-average. This equivalence means that the emergent dynamics of an ergodic system over time can be deduced by studying a cross-section of the phase space at a particular time.

Consider an ergodic system $S^i$ with configuration $i$, where some emergent property of interest has value $y^i$. Now, assume that apparatus $A$ can be used to make indirect measurements of $S^i$ such that $A(S^i) = x_j^i$, and further that this measurement procedure can be modelled via Monte Carlo sampling to simulate a set of $n$ indirect measurements $x^i = \{x_1^i, \ldots, x_n^i\}$. We call these measurements "indirect" because they do not probe the ergodic process directly but instead capture its effects on some other observable property. Returning to the example of temperature, the average kinetic energy of the particles of a system is not measured by observing individual particles directly but by macroscopic effects such as volume changes in an interacting fluid. By the ergodic hypothesis, emergent value $y^i$ can be deduced by studying an appropriate cross-section of the phase space of $S^i$. However, since we do not have direct access to this cross-section, we must instead learn a mapping from indirect measurements of microstates to emergent values. In the absence of complete analytic descriptions, we suggest that such a mapping can nonetheless be learned using neural networks. This task is achieved using a corpus of $m$ training examples of the form $\{(x^1, y^1), \ldots, (x^m, y^m)\}$ and formulated as a problem of manifold approximation (Fig. 2).

Suppose there exist diffeomorphic functions $f$ and $g$ such that $f$ projects every measurement $x_j^i$ onto some smooth manifold $X$ and $g$ projects every emergent value $y^i$ onto some smooth manifold $Y$. Without loss of generality, here we consider the case where every measurement $x_j^i$ is a real-valued vector with dimensions $d$ and every emergent value $y^i$ is a real number. Hence we have that $X$ and $Y$ are embedded in the ambient spaces $R^d$ and $R$ respectively. Now, some of the information contained in each measurement $x_j^i$ may not be useful or necessary to compute $y^i$ and the range of observed emergent values will lie within some bound of $R$. We wish to take advantage of these implicit constraints by learning embedded spaces $X < R^d$ and $Y < R$ as well as a diffeomorphic mapping $\varphi$ between $X$ and $Y$. Formulating the problem in this manner yields the composite transformation $\hat{y}^i = g^{-1} \cdot \varphi \cdot f(x_j^i)$ on the joint manifold $M_{X \times Y} = X \times Y$. The task is to approximate manifolds and mappings that produce $\hat{y}^i$ close to $y^i$. Importantly, since the desired composite transformation is a continuously differentiable function on a compact subset of $R$, it should theoretically be learnable by the universal approximation theorem of neural networks (7).

Consider neural network $N$ with parameters $\theta$, such that $N(x_j^i; \theta) = \hat{y}^i$. Then $N$ can be trained via stochastic gradient descent on the corpus $\{(x^1, y^1), \ldots, (x^m, y^m)\}$ to determine optimal parameters: $\bar{\theta} = \underset{\theta}{\arg\min} \, L(\hat{y}^i, y^i)$, where $L$ is a loss function that captures differences between $\hat{y}^i$ and $y^i$. Importantly, the neural network must learn a mapping that produces the same emergent value across different measurements. That is, $N$ must project measurements corresponding to different points in the ambient coordinate system $R^d \times R$ to the *same* point in the intrinsic coordinate system of $M_{X \times Y}$. This data topology arises because each emergent value corresponds to a distribution of measurements. Hence, for each $(x^i, y^i)$, we consider a neighborhood of points encompassing $(x_1^i, y^i)$, $(x_2^i, y^i)$, …, $(x_n^i, y^i)$ in $R^d \times R$ and the mapping $\sigma = (f, g^{-1})$ between each $(x_j^i, y^i)$ and $(z, \hat{y}^i)$, where $z = f(x_j^i)$ and $\hat{y}^i = g^{-1} \cdot \varphi(z)$. Then $\sigma: R^d \times R \to M_{X \times Y}$ maps to the same coordinate $(x_j^i, y^i) \to (z, \hat{y}^i)$ for all $j = 1, \ldots, n$. Topologically speaking, we say $\sigma$ defines a local coordinate chart of $M_{X \times Y}$ near the neighbourhood of $(x^i, y^i)$. Learning this projection mapping can be thought of as extracting the key features across every measurement in $\{x_1^i, \ldots, x_n^i\}$ that are sufficient to compute $\hat{y}^i$. By producing these low-dimensional feature representations of each $x_j^i$ we have $X < R^d$. Additionally, since the emergent values $y^1, \ldots, y^m$ will span some finite range of $R$ we have $Y < R$. Lastly, defining $\bar{\theta}$ in the manner described above encourages mappings that produce $\hat{y}^i$ close to $y^i$.

To demonstrate the use of our proposed framework we consider an ergodic process in ethanol-water mixtures, namely, proton exchange between water and the hydroxyl group of ethanol which occurs at a rate that depends on the relative concentration of the two compounds. This process is ergodic because the exchange rate associated with a typical exchange-pair over time is equivalent to that at a particular time for an ensemble of pairs. Nuclear magnetic resonance (NMR) spectroscopy can be used to measure the exchange rate through the effects exchange has on the spectrum. In high-field NMR, the Bloch-McConnell equations provide a simple analytical solution to the spectral form in the presence of exchange (8). Since resonance signals originate from individual proton spins, exchange simply represents a swap of resonance frequencies for the two protons, which is incorporated into the semi-classical differential equations describing the precessing spins. These can also be adapted to measure exchange via 2D NMR spectroscopy (9). However, high-field NMR requires large, expensive superconducting magnets. We are developing novel methods to reduce the cost and footprint of NMR by using field strengths orders of magnitude lower. At such low fields, conventional NMR techniques do not work, and we instead utilize homonuclear J-coupling spectroscopy based on the J-synchronized echo (SyncE) sequence (10, 11). Modeling exchange in this regime presents three challenges. First, J-coupling spectroscopy measures magnetic states of strongly-coupled spins grouped together into dressed states, and their energy levels are typically calculated numerically from the Schrodinger equation rather than via perturbation theory. Second, detection occurs during and following a series of pulses which occur during and on the same timescale as the exchange process. Third, there is no known analog to the Bloch-McConnell equations that provides analytic solutions in the strong-coupling regime. Therefore, we instead use Monte Carlo simulation to model J-coupling spectra at various exchange rates via numerical simulation of the system during the SyncE pulse sequence. The DRONE network (12) was trained to determine exchange rates from these simulated J-coupling spectra.

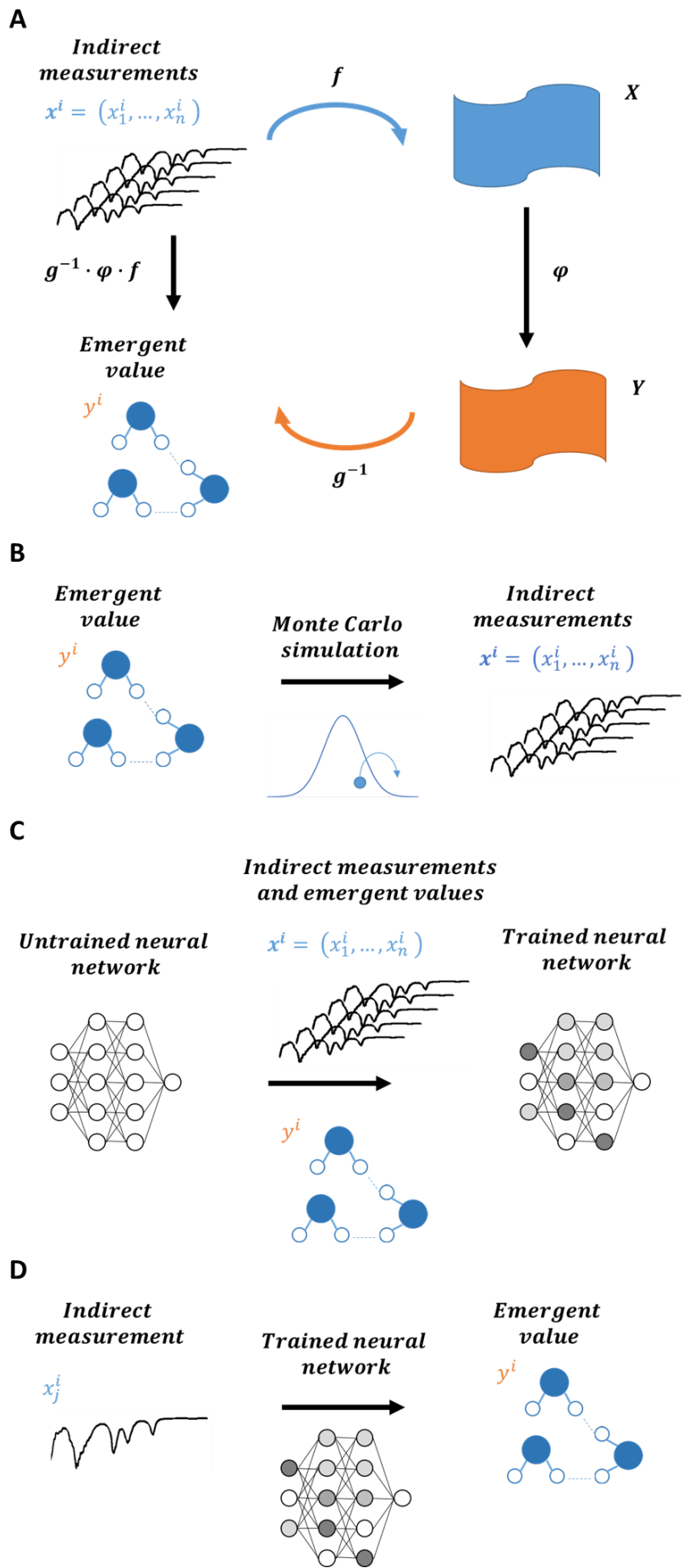

**Fig. 2. Mathematical intuition and workflow for our proposed learning framework** (**A**) The task of mapping from indirect measurements of an ergodic system to emergent properties can be considered as one of manifold approximation, in which $f$ projects every measurement $x_j^i$ onto some smooth manifold $X$, g projects every emergent value $y^i$ onto some smooth manifold $Y$, $\varphi$ is a between-manifold mapping from $X$ to $Y$, and the goal is to estimate emergent values from measurements via the composite transformation $g^{-1} \cdot \varphi \cdot f$. (**B**) A training corpus can be generated via Monte Carlo simulation to produce a set of measurements across a set of emergent values. (**C**) Deep learning can be used to train a neural network to learn a mapping between indirect measurements and emergent values using an appropriate training corpus (**D**) A trained neural network can be deployed to estimate emergent values given indirect measurements.

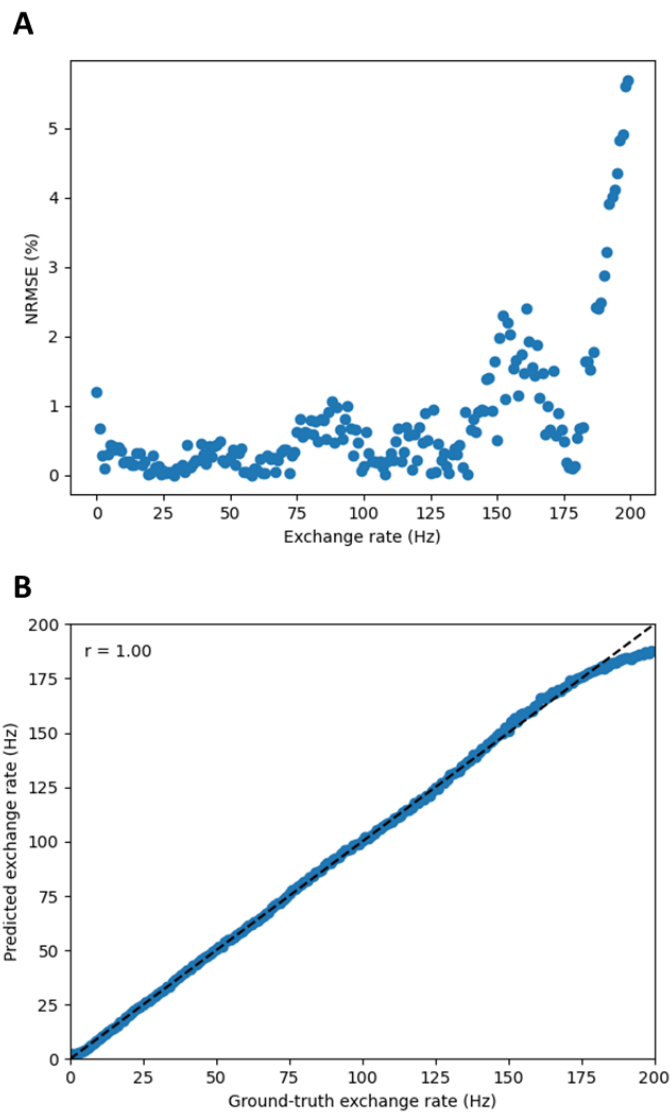

**Fig. 3. Network performance was limited to a specific range of exchange rates** (**A**) Plot of normalized-root-mean-squared-error (NRMSE) against exchange rate in Hz for the trained DRONE network when tested on averaged simulated J-coupling spectra (**B**) Scatter plot and Pearson correlation (r) between predicted and ground-truth exchange rates (with black dotted line representing perfect correlation).

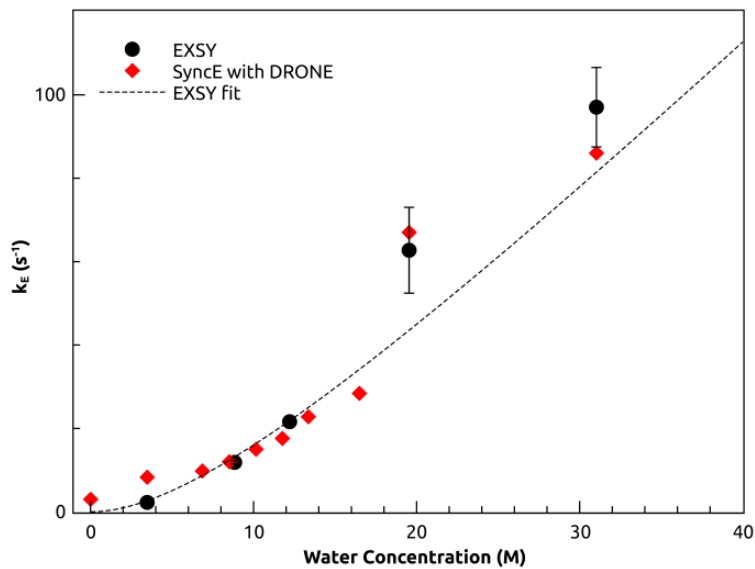

**Fig. 4. Exchange rates measured with SyncE and EXSY** Black dots represent the rate of proton exchange for ethanol, $k_E$, measured with EXSY. Error bars for the first three points are smaller than the symbols. The dashed curve is a best-fit to the EXSY measurements at low water concentrations (< 14 M water). Red diamonds represent $k_E$ determined by SyncE with DRONE. An expected error of 2% from the simulated data gives error bars smaller than the symbols.

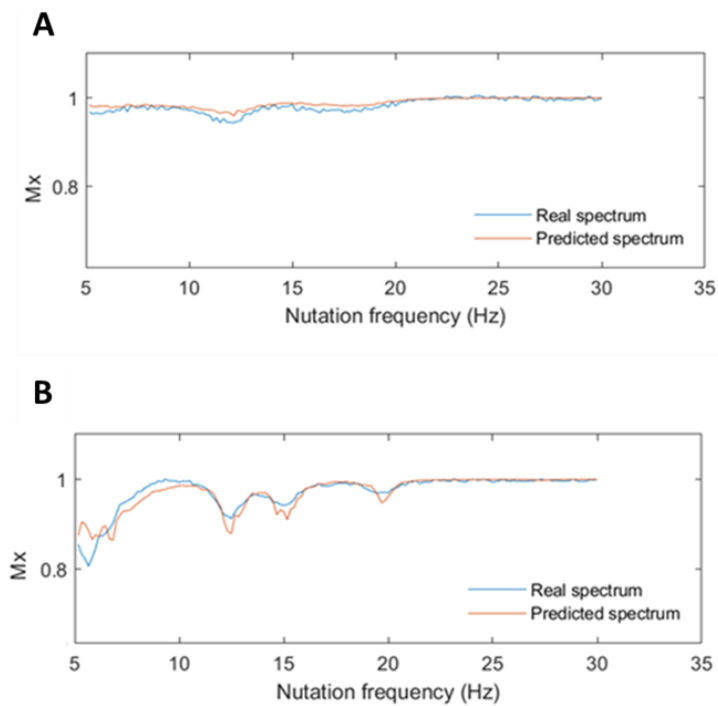

**Fig. 5. Real and predicted SyncE spectra** Measured SyncE spectra were compared with the predicted spectra using Monte Carlo simulation and the DRONE network. Here we show overlapping plots of real and predicted spectra for the best-performing exchange rate at 27.9 s$^{-1}$ (**A**) and the worst-performing exchange rate at 2 s$^{-1}$ (**B**).

## Results

**Performance on data generated by Monte Carlo simulation.** We trained the DRONE network using spectra simulated for exchange rates between 0 and 199 s$^{-1}$ in integer steps, with eight individual spectra provided for each exchange rate. Due to the random timing of exchange events, individual spectra exhibited variations, which were more pronounced at slower exchange rates. This variability enabled robust learning without the introduction of artificial noise. We then tested the trained DRONE network on averaged simulated spectra (Fig. 3). As the system is ergodic, these averaged spectra approximate measurements for an ensemble of exchange pairs. We found good agreement between predictions and the ground-truth for exchange rates between 0 and 150 s$^{-1}$ but network performance decreased for exchange rates beyond this range. This can be seen in the increasing normalized-root-mean-square-error (NRMSE) and deviations in accuracy.

**Performance on experimental data.** We first established an independent measurement of the proton exchange rate using the EXSY sequence in a high-field (600 MHz) NMR spectrometer (Fig. S1). For water concentrations between 0 and 14 M, we found the bidirectional exchange rate as a function of water concentration to be $k = 3.03 \pm 0.04\ s^{-1}M^{-1}$[H$_2$O] (Fig. S2). From this relationship and the relative fractions of exchanging water and ethanol protons, we calculated a predicted curve for the rate of proton loss from ethanol, $k_E$, which is the exchange parameter measured with our low-field technique. EXSY measurements at higher water concentrations deviated from the expect linear relationship. Next, we used DRONE to determine the exchange rate from real data of ethanol-water mixtures acquired with the SyncE sequence at 6.5 mT (276 kHz NMR frequency). For the range 0 to 17 M water, our low-field measurements lie along the EXSY curve. Above 17 M, the data lie above the curve but are still in rough agreement with the values measured with EXSY (Fig. 4). Some of the discrepancies at very low exchange rates are due to self-exchange among ethanol protons, which are captured by SyncE but not EXSY. SyncE measures a self-exchange rate of $k_{EE} = 3\ s^{-1}$ for anhydrous ethanol. We found good agreement between measured and predicted SyncE spectra using DRONE prediction (Fig. 5).

## Discussion

Our proposed learning framework relies on the existence of discoverable relationships between indirect measurements and emergent behavior. At the level of neural networks, these relationships can be understood by probing activations. Where there are detectable differences between two spectra, the activations must also be different. Figure 6 shows activation maps for spectra with exchange rates between 10 s$^{-1}$ and 200 s$^{-1}$. The four simulated spectra for 10 s$^{-1}$ appear to have significant differences due to the natural randomness of the simulation, yet DRONE exhibited identical activation maps, showing that it is able to isolate the important features of the spectra. The spectra at 50 s$^{-1}$ and 100 s$^{-1}$ have different features and activation maps, showing that they are distinguishable from one another. However, 100 s$^{-1}$ and above have nearly identical activation maps, resulting in a decrease in predictive power.

Our measured value of ethanol-water exchange is higher than was found by Luz, Gill, and Meiboom, who measured $k = 0.8\ s^{-1}M^{-1}$[H$_2$O] via lineshape analysis (13). However, we did not attempt to adjust the acidity/basicity of the solutions as they did, and our value is in line with their measurements before they performed adjustments. As in their experiments, it is possible there are some acid or base impurities in the ethanol sample helping catalyze higher exchange rates. The deviation of both the EXSY and SyncE data from the expected linear relationship at high water concentrations also indicates that there are likely higher order exchange processes requiring further investigation.

Learning a mapping between indirect measurements of exchange events and emergent exchange rates required an appropriately curated dataset for training. In ergodic systems such as ours, the average of a series of simulations can be considered as equivalent to the measurement of a large collection of identical systems. Therefore, one possibility is to train the neural network on this averaged measurement for each exchange rate. However, it can take a large number of inputs to produce an averaged result that properly reflects the desired distribution. Moreover, this smoothing may dampen legitimate features that are useful in inferring emergent properties. Alternatively, we show that it is possible to train over a distribution of individual spectra for each exchange rate. Training the network in this manner has four important advantages. First, this training corpus reflects the behavior of ergodic systems, in which a *distribution* over microscopic, stochastic interactions is observed macroscopically as a *single* emergent value. Second, training on individual simulated measurements significantly increases the number of training examples. In the present work, a neural network was trained on 8 simulated acquisitions for each exchange rate rather than the corresponding averages, resulting in an 8-fold enlargement of the training dataset. Third, our previous work on the DRONE network required the addition of Gaussian noise during training to promote robust learning (12, 14), but this was redundant in the present work given the inherent variability of the training data. Fourth, simulated data can be generated abundantly without consuming the resources required for physical experiments. Here we leverage these advantages to learn emergent dynamics from the indirect measurement of ergodic systems. By approximating unknown yet discoverable laws we suggest that this approach could be used to overcome a lack of precise, analytical descriptions in developing novel experimental tools.

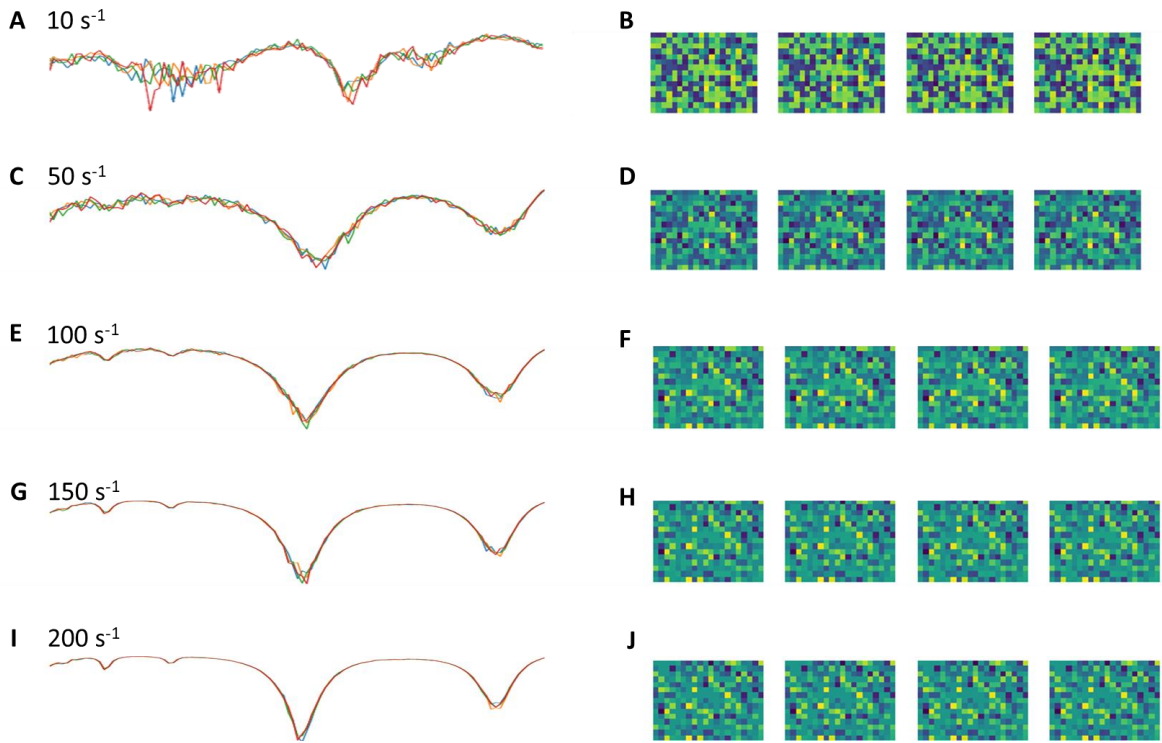

**Fig. 6. Spectra become less distinguishable from one another as exchange rate increases** (**A, C, E, G, I**) The left-hand panel shows J-coupling spectra (zoomed-in for nutation frequencies between 6 and 21 Hz where differences between spectra are most noticeable) across different exchange rates where each simulated measurement is denoted by a different color. (**B, D, F, H, J**) The right-hand panel shows activations for the first connected layer of the DRONE network, where each column corresponds to different input measurements from the exchange rate in each row.

## Materials and Methods

**Experimental Design.** We consider the task of mapping from indirect measurements of an ergodic system to emergent properties as one of manifold approximation via deep learning. This task is achieved using a corpus of $m$ training examples of the form $\{(x^1, y^1), \ldots, (x^m, y^m)\}$, where each is a set of $n$ measurements $x^i = \{x_1^i, \ldots, x_n^i\}$ for emergent value $y^i$ generated by Monte Carlo simulation. A neural network is trained to learn the composite transformation $\hat{y}^i = g^{-1} \cdot \varphi \cdot f(x_j^i)$ on the joint manifold $M_{X \times Y} = X \times Y$, where $f$ projects every measurement $x_j^i$ onto some smooth manifold $X$, g projects every emergent value $y^i$ onto some smooth manifold $Y$, $\varphi$ is a between-manifold mapping from $X$ to $Y$. Neural network $N$ with parameters $\theta$, such that $N(x_j^i; \theta) = \hat{y}^i$, can be used to approximate these manifolds and mappings via stochastic gradient descent to learn optimal parameters $\bar{\theta} = \dfrac{\arg min}{\theta} L(\hat{y}^i, y^i)$, where $L$ is a loss function that captures differences between $\hat{y}^i$ and $y^i$. We validate this framework by using low-field NMR spectra to infer proton exchange rates in water-ethanol mixtures.

Low-field (6.5 mT) J-coupling spectra and high-field exchange spectroscopy (EXSY) measurements were acquired for mixtures prepared from anhydrous ethanol (Sigma Aldrich) and deionized water. These contained between 0% v/v water and 50% v/v water, corresponding to between 0 and 31 M concentrations of water. The DRONE neural network was used to learn a mapping between simulated J-coupling spectra and proton exchange rates. These mapped values from experimentally acquired J-coupling spectra were then compared to exchange rates measured with EXSY.

**Low-field NMR Measurements.** Spectra at 276 kHz (6.5 mT) were measured in a custom-built high-homogeneity electromagnet-based MRI scanner with a Tecmag Redstone™ console described previously (15). For the presently described work, a solenoidal sample coil was used, designed to hold 10 mm NMR tubes, and a B₀ field-frequency lock was used to maintain the resonance frequency within ±0.25 Hz. The scanner was shimmed to achieve a linewidth of deionized water of better than 0.5 Hz. RF pulses directly from the synthesizer were used, resulting in a 90° pulse length of 1 ms using about 4 µW.

Homonuclear J-coupling spectra were acquired using the multi-acquisition J-synchronized echo pulse sequence (SyncE) previously described (16). Echo times $\tau$ varied between 750 ms and 8.3 ms, giving a range of ~0 to 30 Hz for equivalent nutation frequency, $\nu_n$. The actual pulse delays were adjusted appropriately for the pulse width so that the time between the 180° pulse centers was 2$\tau$. We used a total pulse train time $T = 3$ s with $n = 1$ to 179 loops, giving a resolution of 0.167 Hz. All pulses were performed on-resonance with the single $^1$H line of the conventional NMR spectrum. Pulses were calibrated with a Rabi experiment. Echo acquisitions were 8 ms long centered between 180° pulses (16 points with 500 µs dwell time). The delay between measurements was at least 5 T$_1$.

**High-Field NMR.** Exchange spectroscopy was performed at 14.1 T (600 MHz) with a Bruker Bio-Spin Avance NMR spectrometer. Bidirectional exchange rate $k$ was calculated from the intensity of the water and ethanol hydroxyl protons, $I_{WW}$ and $I_{EE}$, respectively, and the cross peaks $I_{EW}$ and $I_{WE}$ following Perrin, et al. (9) and others. Intensities were determined via integration of the appropriate peaks of the 2D NMR spectra. For each mixing time a corrected peak ratio was then calculated taking into account the relative proton concentrations of the two species:

$$r = 4 X_W X_E \left( \frac{I_{WW} + I_{EE}}{I_{WE} + I_{EW}} \right) - (X_W - X_E)^2 \ . \quad (1)$$

Here $X_W$ and $X_E$ are the fraction of exchanging protons coming from water and ethanol, respectively. The value $1/r$ was plotted against mixing time $t_m$ and the data were fit with the function

$$\frac{1}{r} = A \left( \frac{1 - \exp(-kt_m)}{1 + \exp(-kt_m)} \right), \quad (2)$$

where $A$ and $k$ are free parameters found by the fit.

EXSY measures the bidirectional exchange rate $k = k_{EW} + k_{WE}$, where $k_{EW}$ and $k_{WE}$ are the rates of proton exchange from ethanol to water and water to ethanol, respectively. However, SyncE measures $k_E$, the rate of proton loss from ethanol. In the simplest case exchange only takes place with water and $k_E = k_{EW}$. From the mass conservation relationship

$$X_E k_{EW} = X_W k_{WE}, \quad (3)$$

we find

$$k_E = k_{EW} = X_W k. \quad (4)$$

However, we find that SyncE also measures an exchange in the absence of water due to exchange between ethanol molecules, with rate $k_{EE} = 3\ s^{-1}$. The observed exchange rate is then

$$k_E = k_{EE} + k_{EW} = k_{EE} + X_W k \quad (5)$$

Rates $k_{EE}$ and $k$ depend on the molar concentrations of ethanol and water, respectively. Assuming linear relationships, the measurements imply that $k_{EE} = 0.18\ s^{-1} M^{-1}$[EtOH] and $k = 3.03 \pm 0.04\ s^{-1} M^{-1}$[H$_2$O].

**Spectral Simulation.** Custom software was written to efficiently simulate the J-coupling spectrum measured by the multiacquisition SyncE sequence. The software propagates the time-dependent Schrodinger equation for the spin system and calculates the remaining x-axis magnetization, M$_x$, at the center of each echo. To model exchange, we divide the propagation between pulses into a number of steps and calculate the probability $p$ of an exchange event during each step. The probability is

$$p = t_{step} * k_E, \quad (6)$$

where $t_{step}$ is the step length and $k_E$ is the rate of proton loss from ethanol.

For each step, a random number between 0 and 1 is drawn from a uniform distribution, and if the number is less than p, an exchange event occurs. When an exchange occurs, we follow a modified version of the method in Barskiy et al. (17) and replace the hydroxyl proton with one whose spin state is M$_x$. All coherences with the hydroxyl proton are set to zero. (In Barskiy et al., the new hydroxyl proton state is instead assumed to be random, i.e. unpolarized, because their experiment is at zero magnetic field.) As a check, some simulations were also performed with the Spinach package in MATLAB.

**Network Architecture.** The DRONE network (12) was used to predict exchange rates from simulated homonuclear J-coupling spectra. A four layer fully-connected neural network was defined using the Pytorch machine learning framework (18). The input layer consisted of 150 nodes corresponding to the 150 nutation frequencies probed between 5 and 30 Hz during the simulated echo sequences. Data for frequencies below 5 Hz were discarded since they were not useful in determining proton exchange rates. Cropping the data in this manner encouraged the network to focus on the relevant features for inference. As each spectrum was used to compute the corresponding exchange rate, there was only one output node. Between the input and output layers were two hidden layers with 300 nodes each, hence the network required storage of $300 \times 300 = 90,000$ coefficients. A hyperbolic tangent function was used for the hidden layers while sigmoid activation was used for the output layer.

Simulated spectra were generated for ethanol-water mixtures with proton exchange rates between 0 and 199 s-1 in integer steps. Each neural network was then trained using 8 unique simulations for each exchange rate, yielding a training corpus of 1600 spectra and a training:validation split of 90:10 was used. Additionally, each input spectrum was standardized to have zero mean and unit standard deviation while output exchange rates were normalized between 0 and 1. These transformations were used to reflect the shapes of the hyperbolic tangent and sigmoid activation functions, respectively, and were found to drastically improve both training time and validation accuracy. Training proceeded by minimizing the mean squared error between predicted and ground-truth exchange rates using the ADAM optimization algorithm (19) with a learning rate of 0.01 and a batch size of 512. Each network was trained for 50 epochs to ensure convergence, requiring approximately 1 min on a NVIDIA GTX 1080 Ti.

**Acknowledgments:** We would like to thank Mr Danyal Bhutto for organizing computing resources.

**Funding:** NH acknowledges funding support from an Australian Government Research Training Program Stipend, a Cancer Institute New South Wales Translational Program Grant, the Australian-American Fulbright Commission and the Kinghorn Foundation. MSR acknowledges the gracious support of the Kiyomi and Ed Baird MGH Research Scholar Award. This work was partially funded by NSF STTR contract 2014924 (MSR). LLC acknowledges funding support from a NIH Shared Instrumentation Grant S10 OD023406.

**Author contributions:** Conceptualization: NH, SJD, MSR; Methodology: NH, SJD, MSR; Investigation: NH, SJD, EZ, LLC, MSR; Visualization: NH, SJD; Supervision: SJD, MRS; Writing—original draft: NH; Writing—review & editing: NH, SJD, MSR.

**Competing interests:** NH, SJD, MSR have filed a record of invention with Massachusetts General Hospital for the method described in this manuscript.

**Code at:** github.com/ScalarMagnetics/SLIC-Simulator